\documentclass[aps,prl,twocolumn,groupedaddress,showpacs,floatfix]{revtex4}
\usepackage{dcolumn}
\usepackage{amsmath}
\usepackage{amssymb}
\usepackage{graphicx}
\usepackage{bm}
\usepackage[T1]{fontenc}
\usepackage{color}
\usepackage{url}
\usepackage[bf]{subfigure}
\usepackage{rotating}

\begin{document}

\title{Collective behavior in financial market}

\author{Thomas Kau\^{e} Dal'Maso Peron}
\affiliation{Instituto de F\'{\i}sica
de S\~{a}o Carlos, Universidade de S\~{a}o Paulo, Av. Trabalhador
S\~{a}o Carlense 400, Caixa Postal 369, CEP 13560-970, S\~{a}o
Carlos, S\~ao Paulo, Brazil}
\author{Francisco A. Rodrigues}
\email{francisco@icmc.usp.br}
\affiliation{Departamento de Matem\'{a}tica Aplicada e Estat\'{i}stica, Instituto de Ci\^{e}ncias Matem\'{a}ticas e de Computa\c{c}\~{a}o,
Universidade de S\~{a}o Paulo, Caixa Postal 668,13560-970 S\~{a}o Carlos,  S\~ao Paulo, Brazil}

\begin{abstract}
Financial market is an example of complex system, which is characterized by a highly intricate organization and the emergence of collective behavior. In this paper, we quantify this emergent dynamics in the financial market by using concepts of network synchronization. We consider networks constructed by the correlation matrix of asset returns and study the time evolution of the phase coherence among stock prices. It is verified that during financial crisis a synchronous state emerges in the system, defining the market's direction. Furthermore, the paper proposes a statistical regression model able to identify the topological features that mostly influence such an emergence. The coefficients of the proposed model indicate that the average shortest path length is the measurement most related to network synchronization. Therefore, during economic crisis, the stock prices present a similar evolution, which tends to shorten the distances between stocks, indication a collective dynamics.
\end{abstract}

\maketitle

\section{Introduction}

Stock market is an example of complex systems, which are characterized by self-organization and emergent behavior~\cite{Amaral04:EPJ}. The market regulates the relative security prices of companies worldwide without an external or central control.  More specifically, individual actions made by independent investors cause an emergent behavior (the market's direction). Agents, or investors, have information about only a limited number of companies within their portfolio and must follow the regulatory rules of the market and analyze the transactions either independently or in large groupings. Since this complex system is composed of elements that interact nonlinearly generating an intricate organization, it can be naturally represented by networks~\cite{Boccaletti06:PR, Costa07:AP}. In this case, nodes correspond to stocks and the connection between two stocks can be defined according to their price dynamics~\cite{Mantegna99:EPJB, Vandewalle01, Onnela03:PS, Onnela03:PA}. This representation allows studying financial markets by taking into account methods, tools and concepts of complex networks theory~\cite{Newman10, Costa11:AP}.

The early studies involving the representation of financial markets in terms of networks considered the correlation matrix of asset returns~\cite{Mantegna99:EPJB}, in order to find a hierarchical arrangement of stocks through studying the clustering of companies~\cite{Mantegna99:EPJB}. In this way, the initial interest was focused mainly on the topological analysis of financial networks, in which a minimum spanning tree was generated, so as to select the most important connections and group companies according to their asset returns. Subsequent works investigated dynamical aspects of financial networks. For instance, Onnela et. al.~\cite{Onnela03:PRE} studied the resilience of the minimal spanning tree and the consequences of economical events on its structure. They verified that the evolving analysis could capture economic instabilities, such as the \emph{Black Monday}, which occurred on October 19$^{th}$, 1987.

In the current work, we are also interested in examining dynamic aspects of financial market networks. More specifically, it is analyzed the emergence of collective behavior by considering concepts of network synchronization~\cite{Arenas08:PR}. The emergence of collective behavior occurs when stock prices exhibit a similar tendency, defining the market's direction. We have verified that such collective dynamics occurs during financial instabilities. For instance, networks generated during financial crisis (e.g.\ the Black Monday and the the global economic crisis of 2008) present a higher degree of synchronization than networks generated in other periods.

Besides the dynamical analysis, we examine how the network organization changes during financial crisis, which cause the collective behavior. A regression model~\cite{Montgomery} is adopted to verify the network properties that influence the degree of synchronization. It revealed that the increase in the average shortest path length decreased the level of synchronization. Thus, during economic instabilities, since stock prices tend to evolve similarly, as verified by the increase in the network synchronization, the network average shortest path is reduced. The same effect is observed in the average clustering coefficient.  Thus, the emergence of the collective behavior is a consequence of network reorganization during financial crashes.

The following sections discusses the results mentioned above in more detail, describing the construction of financial market networks, the concepts of synchronization and how we can relate the structure and dynamics of these networks by considering regression analysis.

\section{Collective dynamics}

Financial market networks are constructed by considering correlation matrices~\cite{Mantegna99:EPJB}. More specifically, we take into account the correlation coefficient between the time series of two stocks $i$ and $j$, which is given by
\begin{equation}
\rho_{ij}=\frac{\left\langle Y_{i}Y_{j}\right\rangle -\left\langle Y_{i}\right\rangle \left\langle Y_{j}\right\rangle }{\sqrt{(\left\langle Y_{i}^{2}\right\rangle -\left\langle Y_{i}\right\rangle ^{2})(\left\langle Y_{j}^{2}\right\rangle -\left\langle Y_{j}\right\rangle ^{2})}}
\end{equation}
where $Y_{i}$ is the return of a stock $i$ given by $Y_{i} = \ln P_{i}(t) - \ln P_{i}(t-1)$ and $P_{i}(t)$ is the closure at day $t$. The distance between two stocks $i$ and $j$ can be calculated by~\cite{Mantegna99:EPJB}
\begin{equation}
d(i,j)=\sqrt{2(1-\rho_{ij})}.
\label{Eq:d}
\end{equation}
This distance is close to zero if two stocks present similar price evolution, i.e.\ correlated time series. On the other hand, two stocks whose prices have contrary tendency have distance close to two. These distances between $N$ stocks form a symmetric $N \times N$ distance matrix $\mathbf{D}$, which is the weighted matrix that represents the topology of the financial market.

We constructed financial networks by taking into account a database formed by the daily prices of 3,799 stocks traded at New York Stock Exchange. These stock prices are available at the \emph{Yahoo!} financial website~\footnote{\emph{http://finance.yahoo.com}}. We have selected $N=348$ stocks from this set, which are the stocks that have historical data from January 1986 to February 2011. From these data, an amount of 6,008 closure prices per stock is generated.

In order to study the collective behavior of the financial market, it is necessary to consider the time evolution of the respective networks. By setting a time window of length $\Delta t=28 $ days and moving this window along time, we can obtain a sequence of networks --- each one describing the market organization inside each window. This window is moved by an amount of $\delta t=1$ day and a new network is obtained after each displacement. More specifically, the first network is constructed from the time series starting at day $t_{1}^{1}=1$ and ending at day $t_{2}^{1}= 28$, the second network from the time series starting at day $t_{1}^{2} = 2$ and ending at day $t_{2}^{2} = 29$, and so on. This process is repeated until the end of the original time series has been reached. Since the whole database corresponds to a time series composed of 6,007 returns per stock, a total of 5,979 networks was achieved. Differently from previous works (e.g.~\cite{Mantegna99:EPJB,Onnela03:PRE}, instead of using the spanning trees, the fully connected weighted networks were taken into account to represent the financial networks. This choice allowed considering the whole information about network topology, since the spanning tree methodology uses only the most important connections, ignoring the majority of the links.

The collective dynamics of complex systems can be studied by taking into account the concepts of network synchronization~\cite{Watts98:Nature}. This dynamic process causes the emergence of collective phenomena when there occurs the onset of synchronization~\cite{Arenas08:PR}. A model to describe the synchronization of a system was proposed by Kuramoto~\cite{Acebron05:PRM}. In complex networks, each oscillator $i$ obeys an equation of motion given by
\begin{equation}
\dot{\theta}_i = \omega_i + \lambda \sum_{i=1}^{N}a_{ij} \sin(\theta_j - \theta_i), \quad i=1,\ldots,N,
\end{equation}
where $\lambda$ is the coupling strength, $\omega_i$ is the natural frequency of oscillator $i$ (generally distributed according to some function $g(w)$), and $a_{ij}$ are the elements of the adjacency matrix $A$, which represent the topology of the complex systems. More specifically, elements $a_{ij} = 1$ if two nodes $i$ and $j$ are connected, and $a_{ij} = 0$, otherwise. Coupling strengths higher than a determined threshold $\lambda_c$ produce the onset of synchronization. As the coupling strength is increased, more and more oscillators present individual phase around the average phase of the whole system and the network settles in the complete synchronized state.

In order to study the synchronization dynamics in financial market networks, we adapted the Kuramoto model~\cite{Nakagawa94} by considering the distance between stocks, $d_{ij}$, which are the elements of the distance matrix $D$ (see Equation~\ref{Eq:d}).
\begin{equation}
\dot{\theta}_i = \omega_i + \lambda \sum_{i=1}^{N} e^{-\alpha d_{ij}} \sin(\theta_j - \theta_i), \quad i=1,\ldots,N.
\end{equation}
We considered a value of $\lambda = 0.1$. Parameter $\alpha$ is a constant and we adopted $\alpha = 5$. The collective dynamics of the whole system can be measured by the macroscopic complex order parameter,
\begin{equation}
r(t) = \left| \frac{1}{N}\sum_{j=1}^{N} e^{i\theta_j(t)}\right|,
\end{equation}
where $0 \leq r(t) \leq 1$ measures the phase coherence of populations. When $r(t) \approx 1$, all nodes oscillate with similar phases

\begin{figure}[!tpb]
\centerline{\includegraphics[width=1\linewidth]{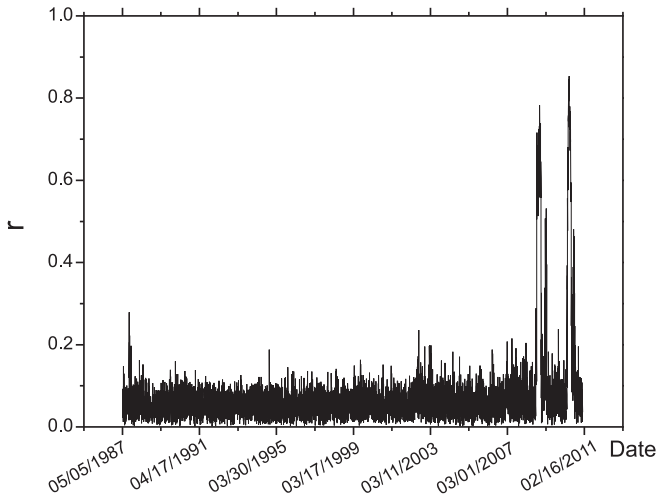}}
   \caption{Time evolution of the network synchronization.}
   \label{Fig:sync}
\end{figure}

Figure~\ref{Fig:sync} presents the evolution of the macroscopic complex order parameter for the generated financial market networks for $t \approx 10^4$. During most of the time, the synchronization level does not suffer a high variation and it is kept around  $r = 0.05$. On the other hand, three prominent peaks can be observed in specific times. The first, at the beginning of the time series, corresponds to the networks constructed between 09/16/1987 and 10/26/1987. A financial instability known as \emph{Black Monday} occurred inside this interval, \emph{i.e.} on October 19$^{th}$, 1987. The other two high peaks correspond to networks generated during the financial global crisis, which started in 2008. The last peak started decreasing only in 2010, when the global crisis was attenuated. Therefore, the time series indicate that during economic instabilities, financial networks tend to become more synchronizable, i.e.\ the phase coherence increases, indicating the emergence of a collective behavior of stock prices, since most of them tend to have a similar evolution. During financial crashes, stock prices move to the same direction presenting a behavior similar to those observed in other type of complex systems, such as cellular movements in tissue formation, flock of birds, and social behavior in human societies. On the other hand, in ``normal'' periods, stock prices tend to evolve in a more independent fashion and the behavior of each stock is not synchronized as verified in financial crisis.

\section{Structure and dynamics}

The synchronization of financial market networks is determined by the network organization. Thus, we investigate the topological properties that influence such synchronization and therefore understanding how financial market networks evolve in different situations. This investigation is one of the main problems in complex networks research~\cite{Boccaletti06:PR}. For instance, Watts and Strogatz~\cite{Watts98:Nature} suggested that the decrease in the average shortest path length in small-world networks facilitated a more efficient coupling and, therefore, enhanced synchronization. They examined this tendency by increasing the probability of rewiring, which creates more shortcuts between pairs of vertices. Other works have investigated the influence of the distribution of connectivity~\cite{Nishikawa03:PRL}, clustering coefficient (e.g.~\cite{Mcgraw05:PRE}) and degree correlations~\cite{Motter05:EL}. In all these papers, networks are modified in order to keep some properties constant while others are varied. However, when the parameters characterizing the original network are modified, other network properties  also change, raising difficulties to draw conclusions about the relationship between one single statistical property of network and its synchronization level~\cite{Arenas08:PR}. To overcome this difficulty, we propose here the use of statistical regression analysis~\cite{Montgomery}. This methodology helps understand how the value of the dependent variable changes when any independent variable is varied --- while the other independent variables are held fixed. Thus, regression allows verifying the relationship between synchronization and network properties in a proper fashion.

Although a regression model does not imply in a cause-effect relationship between variables, it allows to obtain an association between variables. In the case of complex networks theory, it is known that the topology have a direct effect on dynamical process~\cite{Boccaletti06:PR}. So, the cause-effect relationship is determined theoretically and the regression allows to quantify such a relationship.

Three network measurements are considered to determine those that influence the network synchronization. The weight of the connection between two stocks is given by their distance (see Equation~\ref{Eq:d}). The strength of a node, $s$, is defined as the sum of the weights of its corresponding edges. From this measurement, it is possible to quantify the network strength heterogeneity by taking into account the Shannon entropy of the strength distribution, $P(s)$, \emph{i.e.}
\begin{equation}
H = - \sum _{i} P(s_i) \log_2 P(s_i).
\end{equation}\label{eq:entr_degree}
The clustering coefficient for weighted networks is given by~\cite{Onnela05:PRE}
\begin{equation}
cc_{i}^{w} = \frac{1}{k_{i}(k_{i}-1)} \sum_{j,k} \left( \hat{w}_{ij} \hat{w}_{ik} \hat{w}_{jk} \right)^{1/3},
\label{eq:cluster_on}
\end{equation}
where the weights are normalized by the maximum weight in the network, $\hat{w}_{ij} = w_{ij}/\max_{pq}(w_{pq})$, $p,q = 1,2,\ldots,N$. The respective global measurement is the average cluster coefficient
\begin{equation}
C = \frac{1}{N}\sum_{i} ^{N} cc_{i}^{w}.
\label{eq:cluster_average}
\end{equation}
The weighted average shortest path length permits quantifying the average topological distance between the stocks,
\begin{equation}
\ell=\frac{1}{N(N-1)}\sum_{i\neq j}\tau_{ij}.
\label{eq:shortest_path}
\end{equation}
where $\tau_{ij}$ is the length of the shortest distance between stocks $i$ and $j$. Note that a more similar time evolution of a pair of stock prices implies shorter distance between them.

Considering these measurements, we have proposed a regression model to determine the relationship between the structure and dynamics of financial markets. This model is given by
\begin{equation}\label{Eq:model1}
r = \beta_0 + \beta_1 H + \beta_2 C + \beta_3 \ell + \varepsilon.
\end{equation}
where $\varepsilon$ is assumed to be normally distributed with mean zero and standard deviation $\sigma$. This variable captures all other factors which influence the dependent variable $r$ other than the network measurements considered. Thus, in the proposed regression model, the dependent variable $r$ represents the order parameter $r(t)$ obtained for a large value of $t \approx 10^4$.  On the other hand, the independent variables are the network topological measurements, i.e.\ the Shannon entropy of the strength distribution ($H$), the average clustering coefficient ($C$), and the average shortest path length ($\ell$). The contribution of each network measurement $i$ for the synchronization is given by coefficient $\beta_i$.

\begin{figure}[!t]
\centerline{\includegraphics[width=1\linewidth]{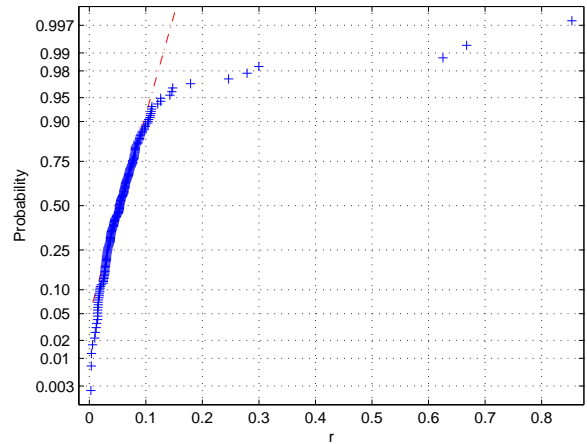}}
   \caption{Normal probability plot for $r$.}
   \label{Fig:normal}
\end{figure}

The regression analysis adopted here considers the least-squares estimator. In this case, some assumptions with respect to the database must be done. First, the networks considered must be independent. Thus, we take into account only networks constructed by disjoint time series, i.e.\ time series without time intersections. More specifically, networks constructed from windows starting at day 1, day 29, day 57, and so on. The second assumption is related to the fact that the independent variable must present, at least approximately, a normal distribution. Figure~\ref{Fig:normal} shows the normal probability plot, which is a graphic technique for normality testing. Despite some outliers, the points lie close to a straight line, which allows concluding that the distribution of $r$ is consistent with a sample from a normal distribution. Ifnteresting to note that the outliers correspond to the networks obtained during financial crisis, i.e.\ they correspond to the highest values of $r$ (see Figure~\ref{Fig:sync}). Thus, during financial instabilities, a financial network has an anomalous organization, differing strongly from normal periods. The presence of these outliers makes the least squares estimation inefficient and can also be biased. Therefore, we adopted the robust regression~\cite{Holland77, Street88} to obtain the coefficients of the regression model, since this method allows treating data with outliers. Robust regression works by assigning a weight to each data point. At the first step, equal weights are assigned to each point and ordinary least squares are used for estimation. In the next step, weights are recalculated so that points farther from predictions in the previous iteration are given lower weight. The coefficients of the model are recalculated by using ordinary least squares. The process is repeated until the values of the coefficient estimates have converged within a specified tolerance.

By performing the robust regression with ordinary least squares, the following linear model was obtained,
\begin{equation}\label{Eq:model2}
r = 1.74 -0.090 H -0.103 C -0.826 \ell + \varepsilon.
\end{equation}
Coefficients $\beta_1=-0.09$ suggests that the value of the Shannon entropy of the strength distribution does not influence significantly the network synchronization. Thus, the heterogeneity in the strength distribution causes a small decrease in the synchronization level. The clustering coefficient and the average shortest path length also contribute negatively to the onset synchronization. A large clustering coefficient implies many transitive connections and, consequently, more redundant paths in the network, which tend to prejudice the emergent behavior. The major influence on the degree of synchronization is due to the average shortest path length, $\beta_3 = -0.826 $. This network property has a negative influence on the network synchronization, since larger distances delay the appearance of the fully synchronized state. During financial crisis, the average shortest path diminishes, leading to an increase int the network synchronization level. This event indicates that the stock prices present a collective behavior, which implies an increase in the correlation between pairs of nodes. During economic crisis, the sock prices tend to exhibit a similar evolution. This result was expected, since the prices tend to suffer a strong decrease during crashes, defining the market's direction. Thus, the increasing in the degree of synchronization is mainly due to the decrease in the average shortest path length during financial crisis.

\section{Conclusions}

Financial markets are examples of complex systems which can be represented as networks. The emergence of a collective behavior has been observed in many complex systems, such as flocking of birds and swarm intelligence in ant colonies~\cite{Mitchell}. This phenomenon is a consequence of the agents which interact according to local rules. Since financial markets are examples of complex systems, a collective behavior is expected to be also observed. The current work analyzes financial complex networks whose nodes represent stocks and the connections between two nodes are established according to a measurement related to the correlation between the temporal price evolutions of the respective stocks. We take into account the Kuramoto model, which describes the synchronization of a system, to study the emergence of collective dynamics. Our results suggest that during financial instabilities, stock prices tend to move to the same direction, becoming more synchronizable, which indicates the emergence of a collective behavior. This tendency was observed during the Black Monday on October 19$^{th}$, 1987, and during the last global crisis, which started in 2008 and was attenuated in the early 2010. In these periods, financial networks presented the highest degrees of synchronization.

In order to determine the topological factors that influence such a synchronization and understand the reorganization of the financial market networks during financial instabilities, a statistical regression model was used. Since the data considered here present some outliers, which correspond to the networks generated during financial crashes, we have taken into account the robust regression approach. The regression model suggests that the clustering coefficient and the average shortest paths length tend to prejudice the network synchronization. Thus, during economic crisis, the stock prices present similar evolution, which results in a higher correlation between time series. Such a correlation tends shorten distances in the networks, enhancing synchronization.

In summary, the results presented in the current work help to understand how the collective dynamics emerge in financial markets. In addition, the regression analysis permits determining how structural properties influence this network dynamics.

\section*{Acknowledgments}
Francisco A. Rodrigues would like to acknowledge CNPq (305940/2010-4) and FAPESP (2010/19440-2) for the financial support given to this research. Thomas K. D. M. Peron would like to acknowledge Fapesp for sponsorship. 

\bibliographystyle{apsrev}
\bibliography{paper}

\end{document}